**Direct Hydrogen Production from Water/Seawater by Irradiation/Vibration-Activated Using Defective Ferroelectric BaTiO$_{3-x}$ Nanoparticles**

Yue Jiang[1*], Cui Ying Toe[2], Sajjad S. Mofarah[1], Claudio Cazorla[3], Shery L.Y. Chang[1,5], Yanting Yin[4], Qi Zhang[1], Sean Lim[5], Yin Yao[5], Ruoming Tian[5], Yuan Wang[6], Tasmia Zaman[1], Hamidreza Arandiyan[6,7], Gunther G. Andersson[4], Jason Scott[2], Pramod Koshy[1], Danyang Wang[1*] and Charles C. Sorrell[1*]

1. School of Materials Science and Engineering, UNSW Sydney, Sydney, NSW  2052, Australia
2. Particles and Catalysis Research Group, School of Chemical Engineering, UNSW Sydney, Sydney, NSW  2052, Australia
3. Departament de Física, Universitat Politècnica de Catalunya, Campus Nord B4-B5, Barcelona 08034, Spain.
4. Flinders Institute for Nanoscale Science and Technology, Flinders University, Adelaide, South Australia 5042, Australia
5. Electron Microscope Unit, Mark Wainwright Analytical Centre, UNSW Sydney, Sydney, NSW  2052, Australia
6. Centre for Advanced Materials and Industrial Chemistry (CAMIC), RMIT University, Melbourne, VIC  3000, Australia
7. Laboratory of Advanced Catalysis for Sustainability, School of Chemistry, University of Sydney, Sydney, NSW  2006, Australia

*Corresponding author:
yue.jiang2@unsw.edu.au; dy.wang@unsw.edu.au; c.sorrell@unsw.edu.au

**Abstract**

Hydrogen is a promising fossil-fuel alternative fuel owing to its environmentally neutral emissions and high energy density.  However, the need for purified water and external power are critical hindrances to implementation of hydrogen production.  The present work reveals the potential to overcome these shortcomings through piezo-photocatalysis of seawater using BaTiO$_{3-x}$ (BTO) nanoparticles.  This material was made piezoelectrically active by annealing under different atmospheres, including O$_2$, N$_2$, Ar, and H$_2$, the latter of which caused Ti$^{4+}$ → Ti$^{(4-x)+}$ multiple reductions and structural expansions that stabilized piezoelectric tetragonal BTO domains.  The resultant defect equilibria combine ionic and electron effects, including Ti redox reactions, charge-compensating surface oxygen vacancy formation, and color centre alterations.  Further, variety of experimental techniques revealed the effects of reduction on the energy band structure.  A strong piezoelectric effect and the presence of self-polarization were confirmed by piezoresponse force microscopy, while simulation work clarified the role of vibration on band bending deriving from the former.  The performance data contrasted H$_2$ evolution using deionized (DI) water, simulated seawater, and natural seawater subjected to photocatalysis, piezocatalysis, and piezo-photocatalysis.  An efficient H$_2$ evolution rate of 132.4 μmol/g/h was achieved from DI water using piezo-photocatalysis for 5 h.  In contrast, piezocatalysis for 2 h followed by piezo-photocatalysis for 3 h resulted in H$_2$ evolution rates of 100.7 μmol/g/h for DI water, 63.4 μmol/g/h for simulated seawater, and 48.7 μmol/g/h for natural seawater.  This work provides potential new strategies for large-scale green H$_2$ production using abundant natural resources with conventional piezoelectric material while leveraging the effects of ions dissolved in seawater.

## 1. Introduction

A hydrogen economy is receiving greater attention as a potential means of dealing with climate change and securing a sustainable clean energy resource. Hydrogen as a sustainable energy carrier is a promising alternative to fossil fuels owing to its environmentally neutral emissions and high energy density of 142 MJ/kg, which is approximately six times that of coal and triple that of gasoline [1,2]. At present, most hydrogen generation is from natural gas by steam reforming using the water-gas shift reaction [3]. However, water splitting is the ideal pathway to achieve efficient hydrogen production owing to its cyclic production-oxidation reactions and the storage/transport capacities in both liquid and gas forms. The principal alternative to the steam reforming is electrolysis although it has the disadvantages of requiring external power input, necessity of separation of the $H_2$ produced at the cathode by the hydrogen evolution reaction (HER) and the $O_2$ produced at the anode by the oxygen evolution reaction (OER), the necessity of the use of purified water, and limited freshwater supplies [4]. Although electrolysis exhibits ~80% efficiency, this can be increased with the use of electrocatalysts, photocatalysts, piezocatalysts, or thermocatalysts [5,6]. Although the associated catalytic processes require energy inputs in the laboratory setting, they can be green and sustainable on the industrial scale through the use of solar cells, sunlight, wind/tide, and geothermal/concentrated sunlight, respectively. Further advantage may be gained through the combination of two or even more of these catalytic processes [7,8]. Although countries such as Australia are poised to leverage all of these processes owing to its high sunlight flux, large land mass, and extensive shoreline, it has the disadvantage of limited supplies of freshwater. This situation highlights the importance of developing means of the splitting of seawater [9].

Seawater is the most abundant aqueous feedstock on earth but there are three principal challenges that will determine the feasibility of direct seawater splitting. First, the similarity of the electrochemical potentials of the OER at 1.23 eV and the chlorine evolution reaction (CER) at 1.72 eV raises the possibility of the production of chlorine gas [10]. Second, when seawater is decomposed, it can generate in solution electron scavengers, such as $Na^+$, and hole scavengers, such as $Cl^-$. Although $Na^+$ appears not to be a serious shortcoming [11], the presence of such scavengers is disadvantageous because they can compete with the HER and OER for catalytically generated electrons and holes, respectively [12]. Third, the microorganisms and other organic matter present in seawater could alter the catalytic process, leading to undesirable side-reactions, or foul and deactivate the catalyst. The preceding risks clarify the desirability of using DI water for water splitting but they also highlight the desirability of developing catalysts for green, sustainable, efficient, and reliable seawater splitting.

One approach to hydrogen production is solar-driven seawater splitting using different semiconductor photocatalysts. Simamora *et al.* [13] used Degussa P25 $TiO_2$ and UV irradiation to contrast the photocatalytic $H_2$ evolution rates for DI water (8.5 μmol/g/h) and simulated seawater (3.1 μmol/g/h). They also observed that the addition of a CuO co-catalyst increased the range of observed wavelengths and consequent $H_2$ evolution rate to 15.7 μmol/g/h. As indicated in **Table S1**, all of these rates are relatively low. In a contrasting study, Peng *et al.* [14] prepared $CdS/TiO_2$ nanocomposites for both DI water and simulated seawater splitting and demonstrated respective $H_2$ evolution rates of 306.3 μmol/g/h and 456.6 μmol/g/h. In this work, sacrificial agents (0.1 mol/L $Na_2S$ + 0.1 mol/L $Na_2SO_3$) were added in order to reduce the recombination tendency of electrons and holes and thereby accelerating the HER [15]. Consequently, this combination of anions appears to have achieved the intended aim, thus suggesting that the beneficial effect of $S^{2-}/SO_3^{2-}$ dominated the deleterious effect of $Cl^-$. The summary offered in **Table S1** [13, 16-22] indicates that photocatalytic seawater splitting is impacted by a range of issues, including materials design complexity, necessity of the use of scavengers, and limited quantum efficiencies [7].

A second approach to hydrogen production is via piezocatalysis, although this has been applied only to DI water at present [23]. This relatively new mechanism is based on the principle of converting cyclic mechanical energy into chemical energy. Such mechanical vibration exists in several easily accessed forms, including tides, as a by-product of wind-generated power, and sound/ultrasound systems. In this process, the piezo-potential induced by mechanical vibration causes electronic band bending, which directs the internal charge carrier flow to the catalyst surface, thereby facilitating the catalytic reaction. The band bending also can improve the photo-generated charge carrier separation by shifting the band edges relative to the HER and OER electrochemical potentials, thus improving the photocatalytic activity [24]. Several piezoelectric materials have been explored as promising candidates for piezocatalysis, including $BaTiO_3$ [7,24,25], ZnO [26,27], $Na_{0.5}K_{0.5}NbO_3$ [28], $BiFeO_3$ [29], and $Na_{0.5}Bi_{0.5}TiO_3$ [30,31].

Hong *et al.* [26] prepared ZnO nanofibers by a hydrothermal method and used these to split DI water using unspecified frequency and power level. This work revealed a very high $H_2$ evolution rate (2192.2 µmol/g/h) and a notable efficiency of conversion of mechanical energy to chemical energy of ~18%. You *et al.* [29] synthesized $BiFeO_3$ square nanosheets using a hydrothermal method and obtained a high $H_2$ production rate of 124.1 µmol/g/h using ultrasound at an unspecified frequency and 100 W. Tetragonal $BaTiO_3$ is a well known wide-band-gap ferroelectric material that has been reported to exhibit both photocatalytic and thermocatalytic functions, although rapid charge carrier recombination hinders its performance [32]. An advantage of tetragonal $BaTiO_3$ is that it demonstrates pressure-induced polarization, which allows the establishment of a dynamic built-in electric field during vibration, *viz.*, the piezoelectric polarization field. The polarization allows the electrons and holes to maintain continuous separation, thus preventing recombination and their consequent mobilization on opposite surfaces, thereby making them available for catalytic hydrogen production [32]. Recently, Su *et al.* [25] developed porous and structurally defective $BaTiO_3$ nanoparticles using a hydrothermal method and examined them for DI water splitting. They obtained a high $H_2$ evolution rate of 159 µmol/g/h at 40 kHz frequency and unspecified power, which they attributed to the presence of surface strain, associated enhanced polarization, and high surface area.

A third approach is based on the potential for additive effects by combining the two approaches of photocatalysis and piezocatalyis as piezo-photocatalysis of DI water. However, the literature is contradictory. Wang *et al.* [33] synthesized $ZnSnO_3$ nanowires using a hydrothermal method and examined them for both dye degradation and DI water splitting using photocatalysis, piezocatalysis, and piezo-photocatalysis. The rhodamine B dye (RhB) degradation efficiency using piezo-photocatalysis was approximately 1.53 and 2.30 times higher those photocatalysis and piezocatalysis, respectively. An $H_2$ evolution rate of 3882.5 µmol/g/h was obtained by the synergistic piezo-photocatalysis, which was higher than those for photocatalysis (3453.1 µmol/g/h) and piezocatalysis (3562.2 µmol/g/h). These trends are supported by the work of Xiao *et al.* [34], who prepared Ag-decorated

Na$_{0.5}$Bi$_{0.5}$TiO$_3$-Ba(Ti$_{0.5}$Ni$_{0.5}$)O$_3$ composite nanopowder by self-propagating high-temperature synthesis. A high H$_2$ evolution rate of 450 µmol/g/h was obtained using piezo-photocatalysis, compared to those of photocatalysis (67.75 µmol/g/h) and piezocatalysis (47.71 µmol/g/h). In contrast, the previously mentioned work by Su *et al.* [25] on defective BaTiO$_3$ revealed that piezo-photocatalysis exhibited a lower H$_2$ evolution rate (103 µmol/g/h) than that of piezocatalysis (159 µmol/g/h), although photocatalysis gave nil results. They attributed this to the different charge transfer mechanisms of the two catalyses. That is, piezocatalysis is driven by the spontaneous polarization/depolarization caused by cyclic compressive stress application/release, respectively, thus establishing a surface piezo-potential. Photocatalysis is inactive owing to insufficient redox potential deriving from the charge carriers. With piezo-photocatalysis, the large number of charge carriers, albeit of low redox potential, becomes sufficient to enhance photocatalysis but, at the same time, the charge carriers screen the surface potential, thus hampering the water splitting process.

A survey of the literature indicates there are no studies of the application of ferroelectrics for piezo-photocatalytic splitting of seawater for H$_2$ production. Consequently, the present work presents an initial investigation of *seawater* splitting using the simultaneously applied piezocatalytic and photocatalytic properties of BaTiO$_3$ [53]. The role of structural defects was investigated through annealing under different atmospheres (O$_2$, N$_2$, Ar, H$_2$) in order to impose specific defect chemistries of different extents. A critical outcome of this strategy was that reduction generated unique structural features comprised of tetragonal domains rich in oxygen vacancies. The observed Ti$^{4+}$ → Ti$^{(4-x)+}$ multiple reductions/expansions and charge-compensating oxygen vacancies effectively stabilize these structural features [35,36], which are graded coherently within a more stoichiometric cubic matrix. The role of these tetragonal domains was explored through photocatalysis, piezocatalysis, and piezo-photocatalysis using dye degradation and water splitting. The BaTiO$_3$ nanopowder annealed under H$_2$ showed highly efficient dye degradation of RhB (99% degradation in 30 mins) and H$_2$ evolution rates of 96.9 µmol/g/h (piezocatalysis), nil (photocatalysis), and 132.4 µmol/g/h (piezophotcatalysis). The comparative H$_2$ evolution rates as a function of water type were 100.7 µmol/g/h (DI water), 63.4 µmol/g/h (simulated seawater), and 48.7 µmol/g/h (natural seawater). These data suggest that highly defective and structurally stable BaTiO$_3$ offers the potential for significant outcomes in large-scale green and sustainable freshwater purification and hydrogen production from seawater.

## 2. Results and Discussion

Commercial cubic BaTiO$_3$ (BTO) nanoparticles (<100 nm) were modified to vary the defect concentrations by heat-treatment under O$_2$, N$_2$, Ar, or H$_2$ at 800°C for 12 h and then cooled to room temperature; the respective samples were denoted BTO-O, BTO-N, BTO-A, and BTO-H. In contrast with the first three, a significant change in color from white to dark grey was observed when the BTO was annealed under H$_2$, thus confirming its greater extent of reduction. Schematic details of the synthesis are shown in **Figure 1(a)** and described in detail in the Experimental Methods in the **Supplementary Information**. **Figure 1(b)** shows that the X-ray diffraction XRD diffraction peaks of all samples can be indexed to BaTiO$_3$ crystals [37]. **Figure 1(c)** shows that the representative main (101) peak for BTO-H exhibits a clear shift to a lower angle, thus suggesting lattice expansion from the previously mentioned Ti reduction (charge compensated by oxygen vacancy formation). The XRD Rietveld refinement data for the BTO structure, which is illustrated in **Figure S1(a)**, are shown in **Figure S1(b)**. The XRD data reveal that the BTO (200) peak exhibits a shoulder at the lower angles, which is indicative of the (002) to (002) + (200) peak splitting and associated onset of the cubic → tetragonal BTO phase transformation [37,38].

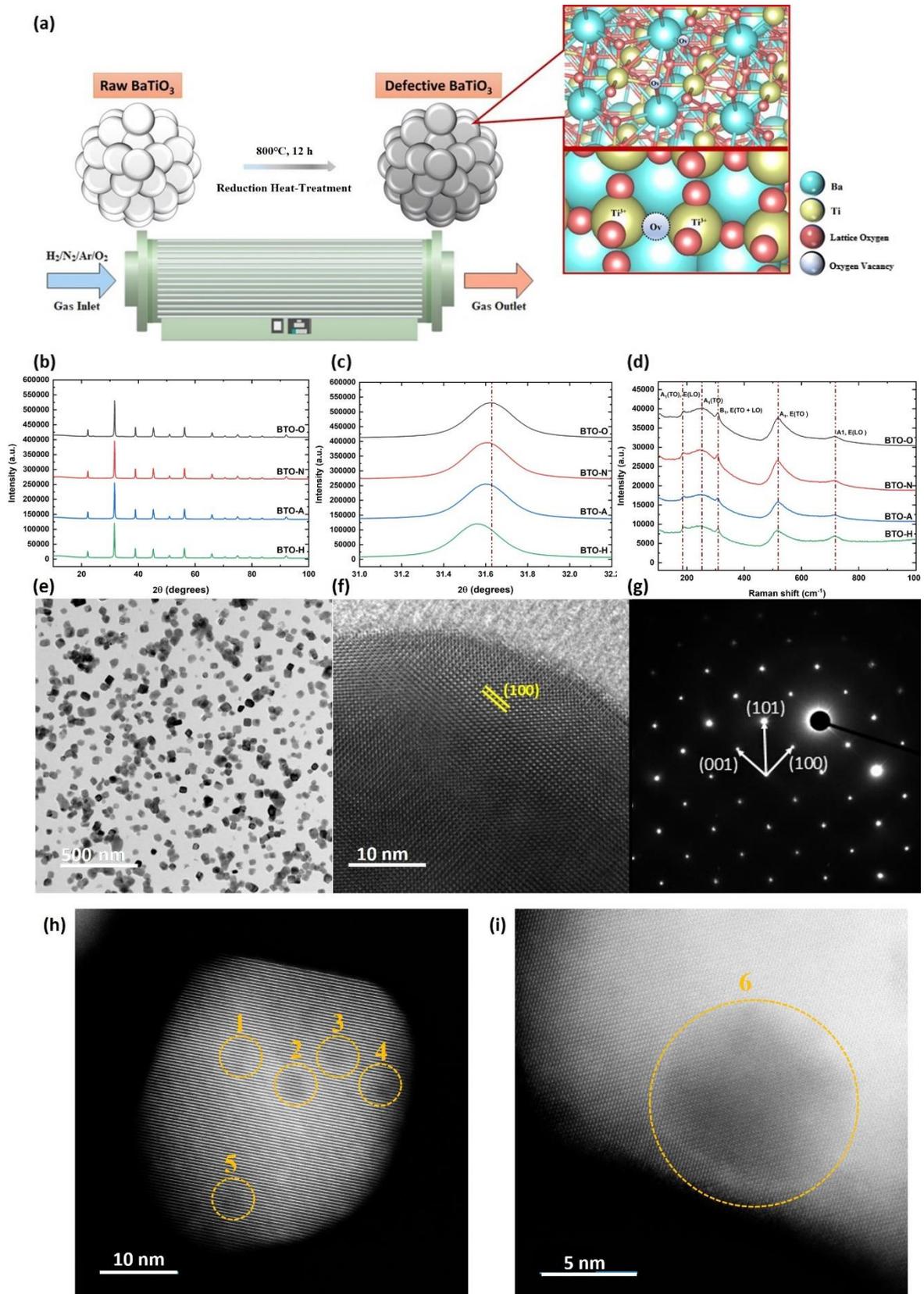

**Figure 1. Synthesis and characterization of defective BTO nanoparticles**: (a) Schematic diagram for synthesis, XRD patterns: (b) range 10°-100° 2θ (identical pattern scales), (c)

range 30°-35° 2θ (identical pattern scales), (d) laser Raman microspectroscopy patterns (identical pattern scales), TEM images of BTO-H: low-magnification bright-field image, (f) high-resolution image, (g) SAED image for BTO-H for (f), HAADF-STEM images of BTO-H: (h) lower magnification image of entire particle (circled regions indicate oxygen-vacancy-rich tetragonal domains), (i) higher magnification image particle edge.

The presence of tetragonal BTO was confirmed by laser Raman microspectroscopy, the spectra for which are shown in **Figure 1(d)**. The main features observed include a small peak at 175 cm$^{-1}$ [$A_1(TO), E(LO)$], a broad peak centered at 265 cm$^{-1}$ [$A_1TO$], a relatively sharp peak at 306 cm$^{-1}$ [$B_1, E(TO + LO)$], a broad sharp peak centered at 520 cm$^{-1}$ [$A_1, E(TO)$], and a small broad peak centered at 720 cm$^{-1}$ [$A_1, E(LO)$] [39]. It is well established that cubic BTO is Raman-inactive owing to the isotropic distribution of electrostatic forces around the centrosymmetric Ti$^{4+}$ ion [37,39,40]. Hence, the appearance of Raman peaks confirms the XRD identification of tetragonal BTO. The Raman-active nature of tetragonal BTO derives from the electronic/ionic polarization caused by the off-centered displacement of the Ti$^{4+}$ cation within the TiO$_6$ octahedron [32], which results in ferroelectric and consequent piezocatalytic behavior [25].

The nanoparticles were examined in greater detail using transmission electron microscopy (TEM). **Figure 1(e)** shows that the BTO-H nanoparticles exhibited monodisperse cuboidal morphologies of typically 30-60 nm size shapes after deagglomeration by 30 min sonication. The lattice fringes shown in **Figure 1(f)** extend throughout the individual grains, thus indicating that they are well-crystallized crystals. The single-crystal nature of the particle shown in **Figure 1(f)**, is demonstrated by the reflections ([001] zone axis) from the selected area electron diffraction (SAED) pattern shown in **Figure 1(g)**. That the grain is highly crystalline is demonstrated by the well defined and bright dots. The tetragonal symmetry is confirmed again by the difference in interplanar spacing of the (001) plane (0.402 nm) and (100) plane (0.400 nm) [41].

The structural homogeneity of BTO-H was examined using high-angle annular dark-field scanning transmission electron microscopy (HAADF-STEM). **Figure 1(h)** and **Figure 1(i)** reveal regions of dark shading that are consistent with enriched oxygen vacancy concentrations. The continuous lattice fringes indicate that these domains are structurally coherent across the graded interface with the lighter matrix. These darker domains, which indicate lower average atomic density, correspond to tetragonal BTO, which is stabilized by the Ti$^{4+} \rightarrow$ Ti$^{(4-x)+}$ multiple reduction/expansion and charge-compensating oxygen vacancies, the dual presence of which is confirmed by the subsequent X-ray photoelectron spectroscopy (XPS) data. This identification is supported by other work that correlates the stabilization of tetragonal BTO with increasing oxygen vacancy concentrations [35,36].

The structural homogeneity was further examined by scanning electron microscopy (SEM), energy dispersive spectroscopy (EDS), laser diffraction, and the Brunaeur-Emmett-Teller (BET) technique. The SEM images of **Figure S3** show that the as-received nanopowders are agglomerated into micron-sized spheres. However, **Figure S1(c)** reveals that sonication for 10 min was sufficient to convert most of the soft agglomerates into the monodisperse particles apparent in **Figure 1(e)**; a small proportion of more stable hard agglomerates was retained. The additional EDS elemental mapping clearly indicates the presence of Ba, Ti, O elements which are uniformly distributed in the nanoparticles. The Brunauer-Emmett-Teller (BET) data for specific surface areas, pore volumes, and average pore sizes are given for the different nanoparticles in **Table S2**. These data do not show clear correlations between the three sets of data, indicating that the macroscopic pore features are not responsible for the progressively increasing trend in surface areas. That is, since the ionic radii of oxygen and nitrogen are within 5% of one another [42], then it is likely that the surface area represents a measure of the access of a nitrogen ion to an oxygen vacancy. Consequently, the surface area represents an alternative means of assessing the oxygen vacancy concentration ($[V_O^{\bullet\bullet}]$). This relation is further illustrated in **Figure S2**, which reveals clear correlation between the surface area and the $[V_O^{\bullet\bullet}]$ determined by XPS (**Table S3**). Further, these data suggest that effectiveness of the different atmospheres in chemical reduction is in the expected order:

BTO-O (Negatively Charged O) < BTO-N (Negatively Charged N) < BTO-A (Neutral Ar) << BTO-H (Positively Charged H)

Since **Figures 1(h)** and **(i)** indicate that the distribution of the oxygen-vacancy-stabilized tetragonal domains is inhomogeneous, then the parallel trends revealed in **Figure S2** suggest that the oxygen vacancies are homogeneously distributed across the surfaces of the grains and that the tetragonal domains are volumetric and are not limited only to the subsurface.

As illustrated in **Figure S1(a)**, which is indicated as a simplified projection in **Figure 2(a)**, BTO has a typical perovskite structure ($ABO_3$), which occurs as both cubic (space group $Pm\bar{3}m$) and tetragonal (space group $P4mm$) polymorphs, where $Ba^{3+}$ occupies the A site and $Ti^{4+}$ occupies the B site. As what effectively is $BaTiO_{3-x}$ has intrinsic oxygen vacancies (assuming ionic charge compensation), the application of $O_2$ probably decreased this concentration while the application of the reducing atmospheres ($N_2$, Ar, and $H_2$) increased it. These atmospheres increase the formation of oxygen vacancies generated by electron migration, *viz.*, polarons [43]. That is, the Ti reduction is caused by electron localization on the Ti ions adjacent to the vacancies, thereby forming a $Ti^{4-x}$-$V_O^{\bullet\bullet}$ dipole [43]. The room-temperature electron paramagnetic resonance (EPR) scan for the defective BTO in **Figure 2(b)** reveals signals at g = 2.002, assigned to surface $V_O^{\bullet\bullet}$, and at g = 1.973, assigned to the $Ti^{3+}$-$V_O^{\bullet\bullet}$-$Ti^{3+}$ complex [24,44,45,46]. The BTO-H, which is the sample reduced by $H_2$, exhibits significantly more intense peaks than those of the other three samples, the difference between which is supported by the trend in **Figure S2**. Further, the intensity of the latter peak is significantly higher than those reported in the literature [44], indicating the relatively high $Ti^{3+}$ concentration ($[Ti^{3+}]$). In order to confirm the assignments of these peaks, the EPR was done at 70°C, as shown in **Figure S4**. While the faster diffusion of the electrons at the higher temperature is consistent with the observed signal attenuation [47], the maintenance of the g values confirms that they do not indicate anomalous phenomena and hence are as assigned. Finally, while the progressive $Ti^{4+} \rightarrow Ti^{(4-x)+}$ reductions are feasible for $H_2$ reduction, no signal for $Ti^{2+}$ and $Ti^0$ is possible because they are EPR-inactive [48].

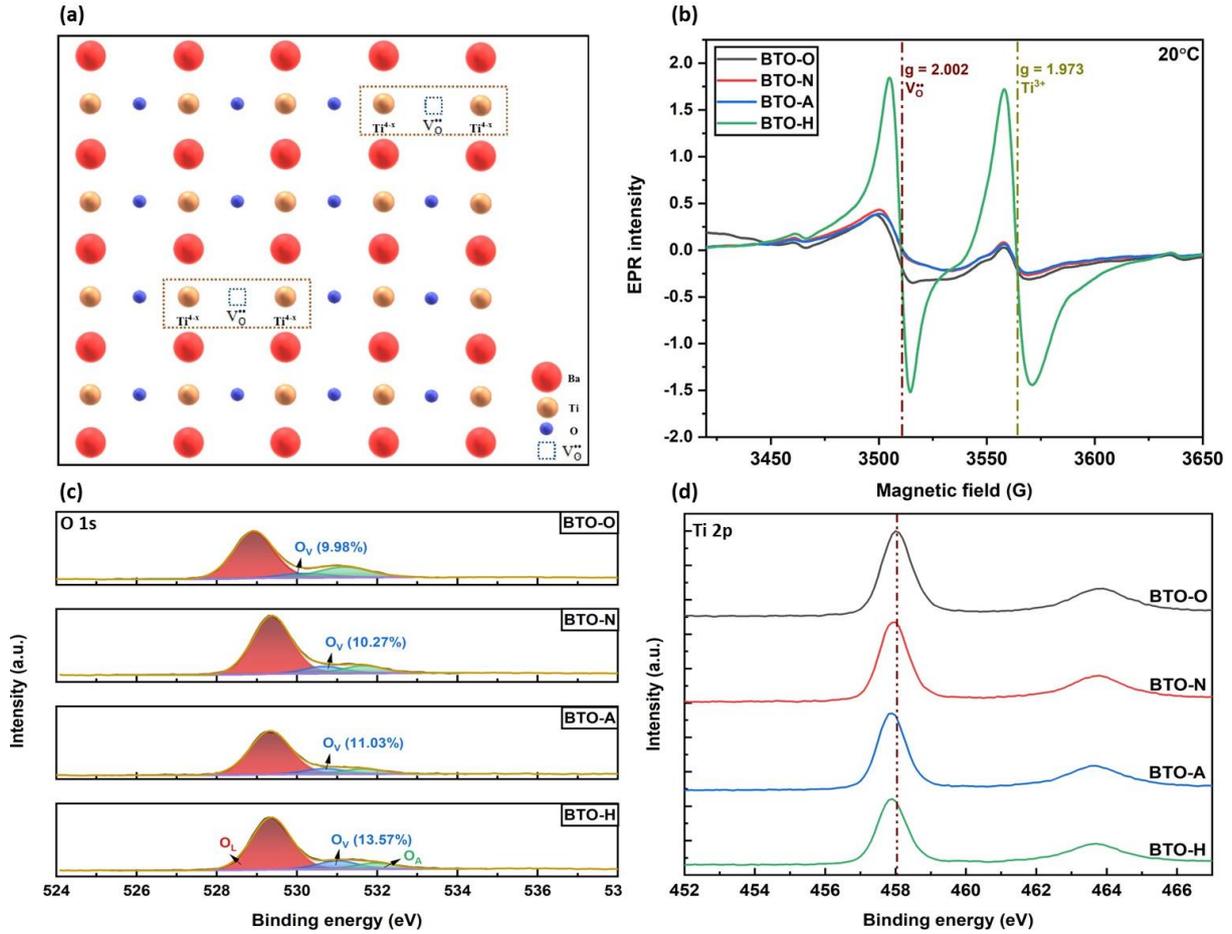

**Figure 2. Defect and chemical analyses of BTO nanoparticles**: (a) Schematic of hypothetical basic defect structure (bulk); (b) Room-temperature EPR spectra (bulk, subsurface, surface); XPS spectra: (c) O 1s (subsurface) and (d) Ti 2p (surface)

**Figures 2(c)** and **2(d)** show the X-ray photoelectron spectroscopy (XPS) spectra for the BTO nanoparticles. The XPS spectra of the O 1s peak for BTO can be deconvoluted into three peaks of 529.6 eV ($O_L$), 530.9 eV ($O_V$), and 531.9 eV ($O_A$), which are ascribed to lattice oxygen, nonlattice ionic oxygen associated with oxygen deficient ($V_O^{\bullet\bullet}$) regions, and adsorbed surface water and/or hydroxyl molecules, respectively [24,33]. Concerning the critical $O_V$ peak, this has been attributed to surface-adsorbed $O^-$ ions, which have a reduced electron density [49]. Consequently, they alter the net electronic charge density and so can be detected by XPS although they often are ascribed nominally to oxygen vacancies [49,50].

As shown in **Figures S2**, **2(b)**, and **2(c)** as well as **Table S3**, the $O_V/(O_L+O_V)$ ratio of 13.57 at% was significantly higher for BTO-H than those of the other samples, *viz.*, BTO-O (9.98 at%), BTO-N (10.27 at%), and BTO-A (11.03 at%), thus confirming the role of thermal reduction in the generation of structural defects in the form of charge-compensating $Ti^{3+}$ (and possibly $Ti^{2+}$) and $V_O^{\bullet\bullet}$ (and possibly the other color centers $V_O^{\bullet}$ and $V_O^{x}$).

The Ba-Ti-O calculated phase diagram shows that the solid solubility homogeneity region for barium titanate in the oxygen-deficient region is bounded by TiO ($Ti^{2+}$) and BaO ($Ba^{2+}$) [51]. Hence, the limiting $[V_O^{\bullet\bullet}]$ is determined by a stoichiometry change from $Ba^{2+}Ti^{4+}O_3$ to $Ba^{2+}Ti^{2+}O_2$, so the maximal $[V_O^{\bullet\bullet}]$ is 33 at%. Thus, the $[V_O^{\bullet\bullet}]$ for BTO-H of 13.57 at% indicates that the extent of reduction was 41% as this is the proportion of the possible oxygen vacancies that could form at the subsurface and in the bulk.

As shown in Figure 2(d), the XPS spectra for the Ti 2p peaks at 458.10 eV (Ti $2p_{3/2}$) and 463.90 eV (Ti $2p_{1/2}$) confirm the dominant presence of $Ti^{4+}$ [52]. The binding energies were obtained by referencing the C1s line to 284.8 eV. Comparison of BTO-O with the other samples reveals that the Ti 2p peaks for the latter shifted to lower binding energies, which is indicative of the less-stable bonding configuration $Ti^{4-x}$-O-$Ti^{4-x}$. Although these shifts are consistent with reduced valences, the shifts are relatively small and within the uncertainty range of the reported binding energy measurements [60]. However, these shifts may indicate minor modification of the $Ti^{4+}$ environment in the form of a small shift in its electron density. Although the most likely cause of this would be an increase in electronegativity upon reduction of $Ti^{4+}$, this is not supported by the values for $Ti^{4+}$ (1.65) and $Ti^{3+}$ (1.5) [92]. Consequently, this shift is consistent with the effect of oxygen vacancies, the absent ions of which have very high electronegativity (3.65).

The depth distribution of structural defects was probed by HAADF-STEM and these data were complemented by determination of the corresponding chemical compositions using electron energy-loss spectroscopy (EELS). **Figure 3(a)** shows the atomic structure of a BTO-H nanoparticle viewed along the [002]/[110] orientation, with the Ba and Ti ionic columns clearly resolved. With the high-collection inner angle used for the HAADF-STEM imaging, the locations of the O ions are not visible [55]. The significantly lower contrast of the Ti ions in the surface relative to those in the subsurface may be attributed to one or both of two causes.

**General Effect:** The structural distortion associated with the surface causes general ionic displacement, as reflected in the projected thickness. This is revealed by the differential contrast between the regions of the surface. Since the major contrast in brightness is for the surface *vs* subsurface/bulk, this indicates that the lattice termination (subsurface) and the bulk are structurally similar and that the surface is more distorted, which are as expected.

**Localised Effect:** The localised distortion associated with the $Ti^{4+} \rightarrow Ti^{(4-x)+}$ reduction causes localised Ti orbital distortion and ionic displacement, as reflected in the differential contrast with the adjacent Ba ions [56]. The latter mechanism is supported by the XPS data. Since the brightness of the Ba ions are similar for the surface, subsurface, and bulk, this indicates the stability of the Ba valence. However, the clear differences in contrast in brightness between the Ba ions and the Ti ions in the surface *vs* subsurface/bulk indicate that there is significant surface $Ti^{4+} \rightarrow Ti^{(4-x)+}$ reduction while this is not the case for the lattice termination (subsurface) or the bulk.

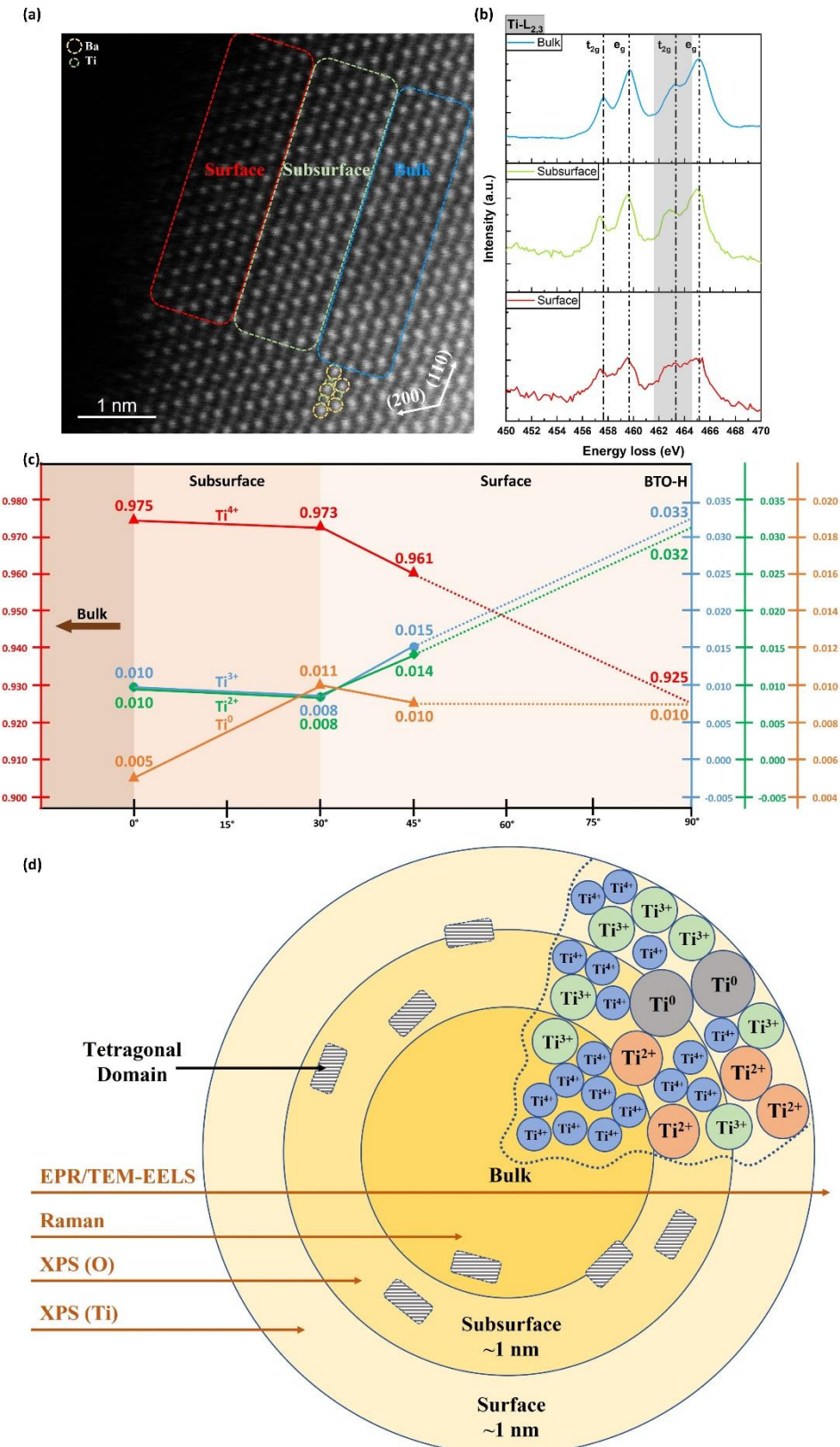

**Figure 3. Analysis of defect concentrations as a function of depth:** (a) HAADF-STEM image of surface, subsurface, and bulk regions of BTO-H; (b) EELS spectra of Ti L edges taken from the surface, subsurface and bulk regions of BTO-H; (c) Normalized ARXPS [$Ti^{4-x}$] for BTO-H extrapolated to 90° (extreme surface); (d) Schematic of graded-reduction and analytical tool penetration depths from surface to subsurface to bulk for BTO-H based on data for (a) to (c)

(10)

**Figure 3(b)** shows the EELS data for the Ti-L$_{2,3}$ edges as a function of depth. The four peaks for the t$_{2g}$ and e$_g$ orbitals correspond to the electronic transitions from Ti 2p$_{3/2}$ (left pair, Ti-L$_3$) and Ti 2p$_{1/2}$ (right pair, Ti-L$_2$) bands [57]. Across the bulk → subsurface, the t$_{2g}$ peak for the Ti-L$_2$ edge shifted to a lower energy loss, which indicates the existence of Ti$^{(4-x)+}$ [25,58]. Further, the peaks become broader from the core to the surface, indicating increasing defect concentrations, particularly charge-compensating oxygen vacancies [45]. Across the subsurface → surface, the principal difference is that the peaks are even broader, thus indicating that there is a significant increase in the defect concentrations across this interface. The observation from EELS is consistent with the angle-resolved X-ray photoelectron spectroscopy (ARXPS), which are discussed immediately following.

**Figure S5(a) and (b)** illustrates the ARXPS data, which were used to examine the valence change in BTO-H as a function of depth. The data for the operating insertion angles of 0°, 30°, and 45° differentiate probing depths by ~2-3 nm. With lower insertion angles, the distances of electrons travelling to the surface toward the detector are increased. Thus, the contributions of electrons from the deeper regions are minimised owing to the electron mean free path [59]. Significantly, the Ti 2p$_{3/2}$ spectra (three left peaks) reveal majority Ti$^{4+}$ (458.7 ± 0.15 eV) and minority Ti$^{3+}$ (456.8 ± 0.15 eV), Ti$^{2+}$ (455.2 ± 0.15 eV), and Ti$^0$ (453.9 ± 0.15 eV) [93-96].

**Figure 3(c)** quantifies these data and extrapolates them to the absolute outer surface. This analysis allows two conclusions. First, in agreement with the EELS data, the inflection at 30° suggests the interface between the surface and subsurface, the former of which is estimated to be of thickness ~1 nm [54]. It is likely that there is a second inflection at a similar distance toward the bulk at a similar distance of ~1 nm (*viz.*, the subsurface thickness), which delineates the bulk/subsurface interface. However, this cannot be detected as the plot terminates by necessity at 0°. Second, the presence of Ti$^0$ is limited effectively to the surface and subsurface. This suggests a potential means of creating a BTO/Ti$^0$ heterojunction structure, which act as electron sinks located on the surface to enhance charge separation. As the concentrations of the other three Ti ions are approximately constant, this suggests that the valence ratios are consistent throughout the bulk, thus demonstrating the high diffusivity of hydrogen in BTO and that the reduction is a bulk effect, not a surface effect.

**Figure 3(d)** shows a schematic that combines the conclusions based on the EPR, EELS, and ARXPS data in terms of the concentration gradients for surface → subsurface → bulk. The presence of the oxygen-vacancy-stabilized tetragonal domains in the subsurface and bulk, the approximate surface and subsurface thicknesses, and the different analytical tool penetration depths also are shown.

The principal reason for the differences in compositions and concentrations between surface, subsurface, and bulk are suggested by the ordered nature of the bulk; the semiordered nature of the subsurface, which is the lattice termination; and the more disordered nature of the surface. In the present case, these effects are revealed by the origins of the Ti$^{4+}$, Ti$^{3+}$, Ti$^{2+}$ and Ti$^0$. Although these ions represent intrinsic defects, they also can be viewed as extrinsic surface defects owing to their origin, which would be one or more of the following:

- Adsorbates from unwashed raw materials
- Unwashed reaction products
- Products of inhomogeneous mixing
- Products of inhomogeneous reaction

- Establishment of concentration gradients during precipitation
- Establishment of concentration gradients during reaction
- Ions forced by stress from the bulk to the surface by segregation
- Chemisorbed ions in the subsurface that have detached from stress/repulsion and become physisorbed in the surface

Since EPR probes through the entire volume and **Figure 2(b)** demonstrates the presence of $Ti^{3+}$, then this valence is confirmed as being present in the bulk. However, the presence of $Ti^{2+}$ and $Ti^0$ in the bulk cannot be probed by EPR as these valences are inactive [48]. These reductions are associated with progressive increases in the ionic radii [42] of $Ti^{4+}$ (0.0745 nm), $Ti^{3+}$ (0.081 nm), and $Ti^{2+}$ (0.100 nm), terminating with the metallic radius of $Ti^0$ (0.147 nm). The approximate doubling of the radius upon full $Ti^{4+} \rightarrow Ti^0$ reduction is likely to the cause of the limitation of the metal to the surface and subsurface.

**Figure 4(a)** shows the photoluminescence (PL) data for all four samples. These data reflect the presence of point defects that generate energy transitions between ground and excited states associated with luminescent centers, which are often color or F centers [61]. The peak intensities are considered to be inversely proportional to the rates of electron-hole recombination, so lower intensities reflect more significant effects, which are controlled by the diffusion distance (*i.e.*,), the presence of trap states (*i.e.*, defects), and the presence of heterojunctions (*i.e.*, interfaces) [54]. The spectra are comprised of two groupings of broad peaks that are attributed to the near-band edge (NBE) and deep-level defect emissions (*e.g.*, oxygen vacancies, surface states, OH⁻, and asymmetry imposed by $Ti^{(4-x)+}$) [62]. These two groupings are comprised of the assignments of five transitions, where the inverse intensities are low for the ionic charge transitions and high for the electronic charge transitions. This bimodal distribution is expected as the amount of energy and time required to break bonds (to form oxygen vacancies) are greater than those to transfer an electron. Hence, the energy and time to create a color centre ($F^0(V_O^{\bullet\bullet}), F^+(V_O^{\bullet}), F^{++}(V_O^x)$) are greater than those to add an electron ($V_O^{\bullet\bullet} \rightarrow V_O^{\bullet}, V_O^{\bullet} \rightarrow V_O^x$). There are no effects from $Ti^0$ because PL is inactive to metals as each exhibits a continuum of electronic states around the Fermi level, so no radiative transitions are possible between states above and below the Fermi level [91].

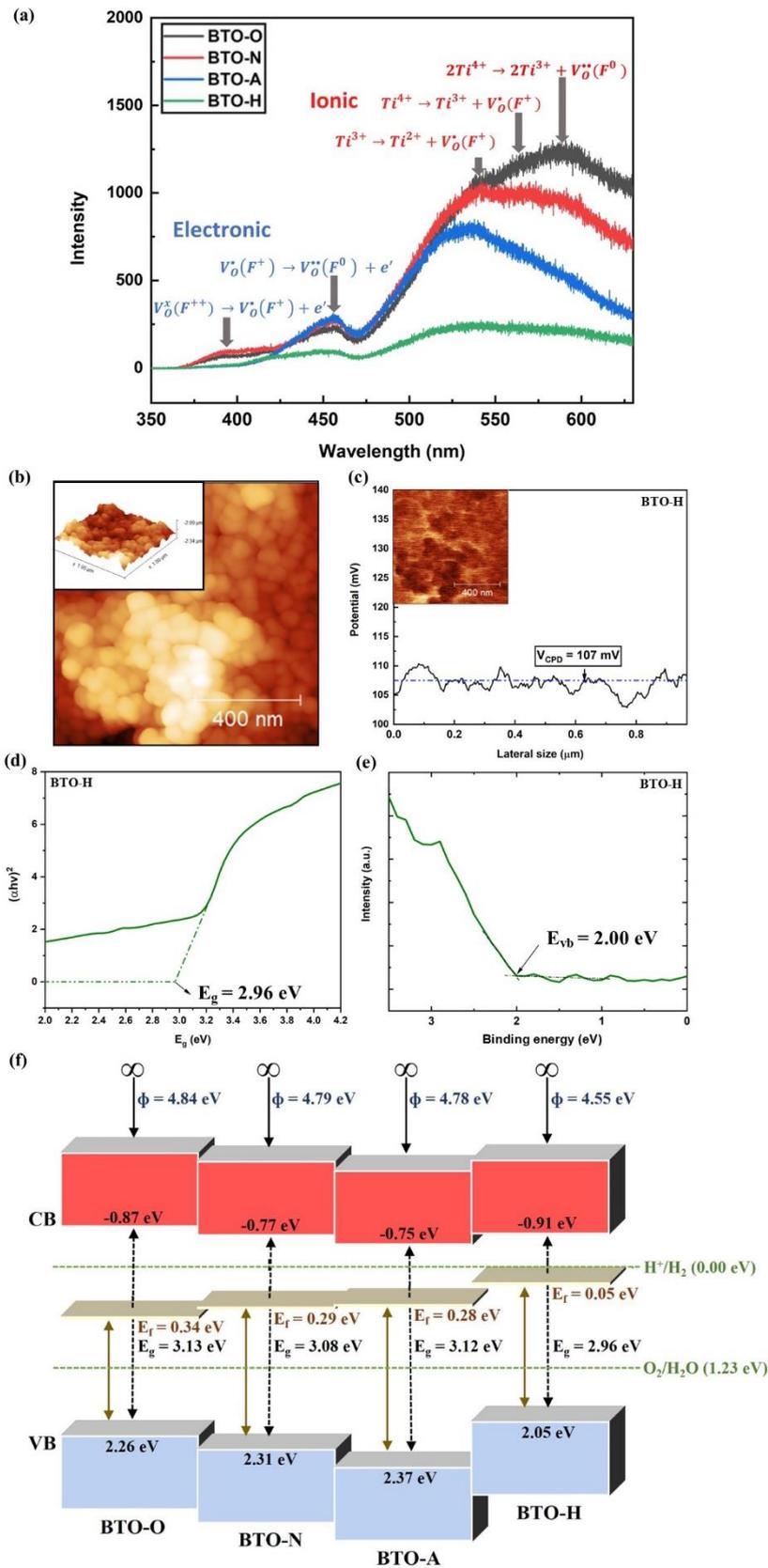

**Figure 4. Band structure characteristics of defective BTO nanoparticles:** (a) PL spectra (identical pattern scales), (b) AFM topography of BTO-H drop-cast film, (c) Contact potential difference (CPD) of BTO-H drop-cast film by KPFM, (d) Kubelka-Munk plots from UV-Vis reflectance spectrophotometry data for optical indirect band gap (e) XPS valence band plots, (f) Resultant energy band diagrams (pH = 0)

(13)

These transitions can be described in terms of Kröger-Vink notation applied to the potential defect equilibria. As the present work focusses on the passive reduction (oxygen partial pressure <21 vol% O$_2$) and active reduction (chemical effect) of BaTiO$_3$ to form $V_O^{\bullet\bullet}$, then the defect equilibria are intrinsic and the possible charge compensation mechanisms for the formation of $V_O^{\bullet\bullet}$ are ionic, electronic, or redox [54,63]. Only Ti$^{3+}$ as the reductant is considered for the sake of simplicity.

Ionic Charge Compensation

$$2Ba_{Ba}^{\times} + 2Ti_{Ti}^{\times} + 6O_O^{\times} \xrightarrow{BaTiO_3} 2Ba_{Ba}^{\times} + 3V_{Ti}'''' + 6V_O^{\bullet\bullet} + 3Ti_S^{\times} + 3O_2 \text{ (g)} \quad (1)$$

Electronic Charge Compensation

$$Ba_{Ba}^{\times} + Ti_{Ti}^{\times} + 3O_O^{\times} \xrightarrow{BaTiO_3} Ba_{Ba}^{\times} + Ti_{Ti}^{\times} + 3V_O^{\bullet\bullet} + 6e' + 1½O_2 \text{ (g)} \quad (2)$$

Redox-Ionic Charge Compensation

$$2Ba_{Ba}^{\times} + 2Ti_{Ti}^{\times} + 6O_O^{\times} \xrightarrow{BaTiO_3} 2Ba_{Ba}^{\times} + 2Ti_{Ti}' + V_O^{\bullet\bullet} + 5O_O^{\times} + ½O_2 \text{ (g)} \quad (3)$$

Redox-Electronic Charge Compensation

$$Ba_{Ba}^{\times} + Ti_{Ti}^{\times} + 3O_O^{\times} \xrightarrow{BaTiO_3} Ba_{Ba}^{\times} + Ti_{Ti}' + h^{\bullet} + 3O_O^{\times} \quad (4)$$

In terms of the formation of $V_O^{\bullet\bullet}$, **Equation 4** is anomalous in that reduction to form $V_O^{\bullet\bullet}$ is absent but its presence would negate the necessity of charge compensation by $h^{\bullet}$. These potential defect equilibria demonstrate that neither ionic (**Equation 1**) nor electronic (**Equation 2**) charge compensation are possible because there is no Ti reduction. In contrast, both of the redox mechanisms are feasible because they include Ti reduction. That is, redox-ionic (**Equation 3**) combines ionic ($V_O^{\bullet\bullet}$) and redox ($Ti_{Ti}'$) charge compensation and redox-electronic (**Equation 4**) combines electronic ($h^{\bullet}$) and redox ($Ti_{Ti}'$) charge compensation.

The formation of oxygen vacancies is unambiguous as this is supported by the XRD, HAADF-STEM, BET, EPR, XPS, EELS, ARXPS, and PL data. Since electronic charge compensation does not involve oxygen vacancy formation, then the mechanism of charge compensation is redox-ionic (**Equation 3**). Thus, the representative defect equilibria representing the remaining reductions from Ti$^{4+}$ in BTO to Ti$^{2+}$ and Ti$^0$ are:

Redox-Ionic Charge Compensation

Ti$^{2+}$: $\quad Ba_{Ba}^{\times} + Ti_{Ti}^{\times} + 3O_O^{\times} \xrightarrow{BaTiO_3} Ba_{Ba}^{\times} + Ti_{Ti}'' + V_O^{\bullet\bullet} + 2O_O^{\times} + ½O_2 \text{ (g)} \quad (5)$

Ti$^0$: $\quad Ba_{Ba}^{\times} + Ti_{Ti}^{\times} + 3O_O^{\times} \xrightarrow{BaTiO_3} Ba_{Ba}^{\times} + Ti_{Ti}'''' + 2V_O^{\bullet\bullet} + O_O^{\times} + O_2 \text{ (g)} \quad (6)$

Redox-Electronic Charge Compensation

Ti$^{2+}$: $\quad Ba_{Ba}^{\times} + Ti_{Ti}^{\times} + 3O_O^{\times} \xrightarrow{BaTiO_3} Ba_{Ba}^{\times} + Ti_{Ti}'' + 2h^{\bullet} + 3O_O^{\times} \quad (7)$

(14)

$Ti^0$:
$$Ba_{Ba}^\times + Ti_{Ti}^\times + 3O_O^\times \xrightarrow{BaTiO_3} Ba_{Ba}^\times + Ti_{Ti}'''' + 4h^\bullet + 3O_O^\times \quad (8)$$

In order to investigate the effects of the atmosphere on the corresponding electronic band structures, XPS was used to determine each gap between valence band (VB) and the Fermi level ($E_f$), UV-Vis spectrophotometry was used to determine the optical indirect band gap ($E_g$), and amplitude-modulated Kelvin probe force microscopy (AM-KPFM) was used to determine the work function (ϕ). **Figure 4(b)** illustrates the topography of the necessary BTO-H drop-cast film, which generated a contact potential difference ($V_{CPD}$) shown in **Figure 4(c)**. The value of 107 mV is significantly higher than that of the other samples (**Figure S6**), indicating enhanced adsorption of the free charges (*i.e.*, piezoelectric potential), which were formed from the external strain, on the polar surface of the nanoparticles [25]. The resultant improved piezoelectric potential facilitates efficient surface reactions.

The CPD is defined as

$$CPD = \Phi_{tip} - \Phi_{sample} \quad (9)$$

where:

$\Phi_{tip}$ = Work function of the metallic tip
$\Phi_{sample}$ = Work function of the sample surface

The work function of Si-coated tips is calibrated using highly ordered pyrolytic graphite (HOPG; $\varphi_{HOPG}$ = 4.6 ± 0.1 eV) [64] to convert the measured CPD to the absolute work function. The absolute surface work function of the sample is calculated as follows:

$$\Phi_{sample} = 4.6\, eV + CPD_{HOPG} - CPD_{sample} \quad (10)$$

where:

$CPD_{HOPG}$ = contact potential difference between the AFM tip and the HOPG reference
$CPD_{sample}$ = contact potential difference between the AFM tip and the sample

Since the measured $CPD_{HOPG}$ and $CPD_{BTO-H}$ were measured by 123 mV (**Figure S6**) and 107 mV (**Figure 4(b)**), respectively, then the work function for BTO-H is calculated as 4.533 eV [64]. The UV-Vis spectra for the optical indirect band gap calculation of BTO-H are shown in **Figure 4(d)**. The Kubelka-Munk method [65] was used to calculate the optical indirect band gap of 2.96 eV using the **Equation 11**:

$$(F(R_\infty)h\nu) = A(h\nu - E_g)^2 \quad (11)$$

Where:

$R_\infty$ = Relative diffuse reflectance
$h$ = Planck's constant
$\nu$ = Frequency
$A$ = Constant
$E_g$ = Optical indirect band gap

The XPS plot for the valence band edge ($E_{VB}$) of BTO-H of 2.00 eV is shown in **Figure 4(e)**.

The preceding data for the $E_g$, $E_{VB}$, and $V_{CPD}$ for BTO-O, BTO-N, and BTO-A are shown in **Figure S6**.

On the basis of these data, the energy band levels were calculated, a schematic of which is shown in **Figure 4(f)**. The key outcomes of these data are that, for the increasing reduction from BTO-O to BTO-H, the $E_g$ decreased from 3.13 eV to 2.96 eV and the $E_f$ rose from 0.34 eV to 0.05 eV, which is in very close proximity to the $H_2/H_2O$ redox potential (0.00 eV) for HER. The former ($E_g$) is attributed to the abundant defect concentrations ($Ti^{3+}$, $Ti^{2+}$, $Ti^0$, $V_O^{\bullet\bullet}$), which are midgap states, that also act as catalytically active sites. When these midgap states are close to the CB (n-type) or close to the VB (p-type), they represent shallow energy levels, thus effectively reducing the $E_g$ and enhancing charge carrier separation by promoting the diffusion of charge carriers to the surface [63]. The latter ($E_f$) is attributed to the effect of F centers. It is known that the $E_f$ can be raised significantly by F centers, even to an energy above that of the conduction band minimum (CBM) [54]. Consequently, the electrons are trapped near the CBM and so they are conducting and have high mobilities. In effect, the proximity of the increased $E_f$ to the redox potential of HER in the defective BTO indicates that the band levels are consistent with effective water splitting.

First-principles calculations based on density functional theory (DFT) simulations were undertaken in order to assess the effects of ultrasound vibration on the energy band gap and band alignments of BTO. Periodic compressive and tensile strains associated resulting from ultrasound vibrations were simulated statically using negative and positive uniaxial strains, η, applied along the polar axis of tetragonal BTO (see **Supplementary Information, DFT Simulations**). The maximal strain amplitude considered in the DFT simulations was 4%. Although this value is likely to be larger than experimentally achieved with ultrasound, it allowed unambiguous determination of the photocatalytic trends induced by uniaxial strain applied to BTO.

**Figure 5(a)** shows the calculated η (strain) dependence of $E_g$. In the absence of strain, the computed value of 2.99 eV is in good agreement with present experimental measurement of 3.12 eV. It can be seen that $E_g$ is only slightly increased under tension but it is reduced significantly under compression. This is interpreted in terms of the effect of direction on the tetragonality, where tension maintains the structure while compression shifts it toward pseudo-cubic. **Figure 5(b)** shows the CB and VB variations estimated for a point located at the center of the simulated BTO slab (see **Supplementary Information, DFT Simulations**) expressed as a function of uniaxial strain. Under tensile strain (η > 0), both the CB and VB levels shift linearly toward lower potentials in a similar manner. These data suggest the presence of an internal electric field, $E_{int}$, which increases with η owing to an increasing accumulation of charges at the opposite surfaces. The accumulation of opposite charges derives from the piezoelectric effect (*i.e.*, $E_{int} = Q/A\varepsilon$, where Q = polarization and piezoelectric surface charge, A = surface area, $\varepsilon$ = dielectric constant) and this produces significant band bending. Under compressive strain (η < 0), the band bending mechanisms are same as under tension except that the potential shifts are of opposite direction (*i.e.*, the CB and VB levels shift towards higher potentials). However, the $E_g$ reduction shown in **Figure 5(a)** for η < 0 reveals differential trends in the CB, where the ascending variation is consistent with electrostatic effect of the internal electric field and the descending variation is consistent with the electronic effect of the $E_g$. Thus, under both compression and tension, the CB approaches the HER potential. In contrast, under compression, the VB approaches the OER potential but, under tension, it

(16)

deviates away from the OER. However, the slope under tension is greater than that under compression, which is an important distinction as the sum of the two strain conditions during amplitude switching must be considered. These differential η-induced variations in the CB and VB potentials are consistent with overall enhancement of the HER by the CB trends, where both strain states cause approaches to the HER potential, and overall hindrance of the OER by the VB trends, where the greater slope under tension dominates the lower slope under compression.

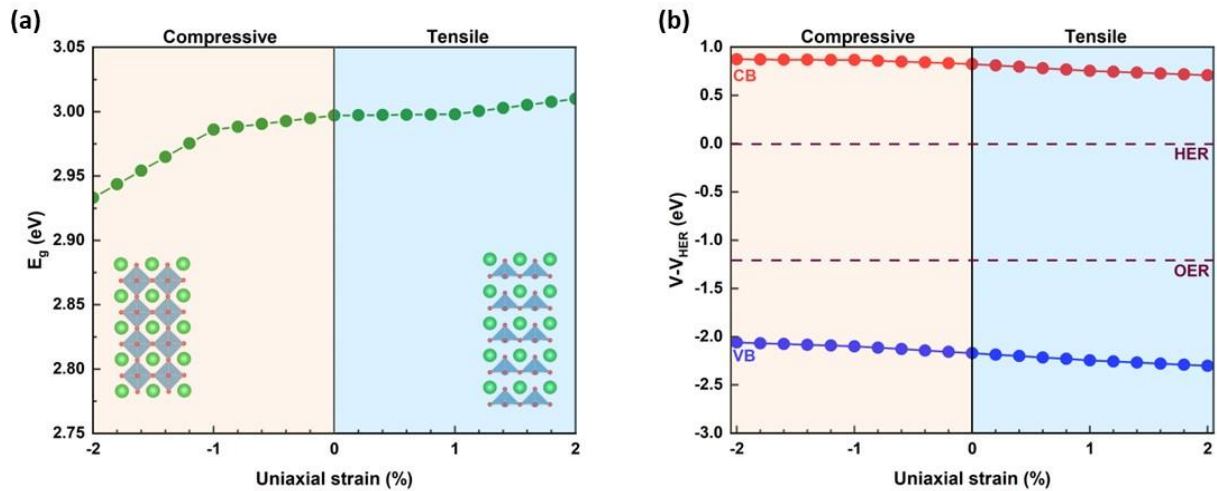

**Figure 5. DFT calculated band structure characteristics of defective BTO nanoparticles:** (a) Band gap change (ordinate scale range = 0.3 eV), (b) Band alignment change (ordinate scale range = 4.0 eV); external uniaxial strain range –2% (compressive, delta c < 0) to +2% (tensile, delta c > 0); strain level is commensurate with probable maximal lattice distortion imposed by ultrasound agitation.

The local piezoelectric properties of defective BTO were characterized by piezoresponse force microscopy (PFM) working in dual AC resonance tracking mode (DART) in order to minimize cross-signalling between the topography, PFM amplitude, and phase. The nanoparticles were deposited on indium tin oxide (ITO)-coated glass substrates by drop-casting. Topography, amplitude, and phase image of the defective BTO-H are shown in **Figures 6(a-c)**, respectively. A nanoparticle with diameter size of approximately 40 nm can be observed with strong PFM amplitude and phase contrast, which is consistent with the results in **Figure 1(e)** and **Figure S1(c)**. The dark contrast in the BTO-H phase image unambiguously indicates a polarization-down state prior to exposure to a strong DC electric field, which is consistent with the observed positive voltage offset shown in **Figure 6(a)**. The PFM amplitude-voltage butterfly curve and phase-voltage hysteresis loop of a typical BTO-H nanoparticle are shown in **Figure 6(d)**. The strong ferroelectric characteristics of the BTO-H are indicated by the nearly 180° phase switching, and the unambiguous piezoresponse is evidenced by the classical PFM amplitude butterfly loop [66]. The asymmetric loops, which have a voltage offset of ~2 V, suggest the presence of significant self-polarization [67]. This often originates from a stress gradient or an internal bias field owing to the nonuniform distribution of oxygen vacancies [68]. **Figures 6(d)** and **S7** reveal that all four samples exhibit such asymmetries and offsets in the PFM butterfly loops but they are more pronounced in BTO-H while those of the other three samples are similar. These results are consistent with and hence are mutually confirmatory of the data in **Figures S2**, **2(b,c)**, and **3(a)**.

(17)

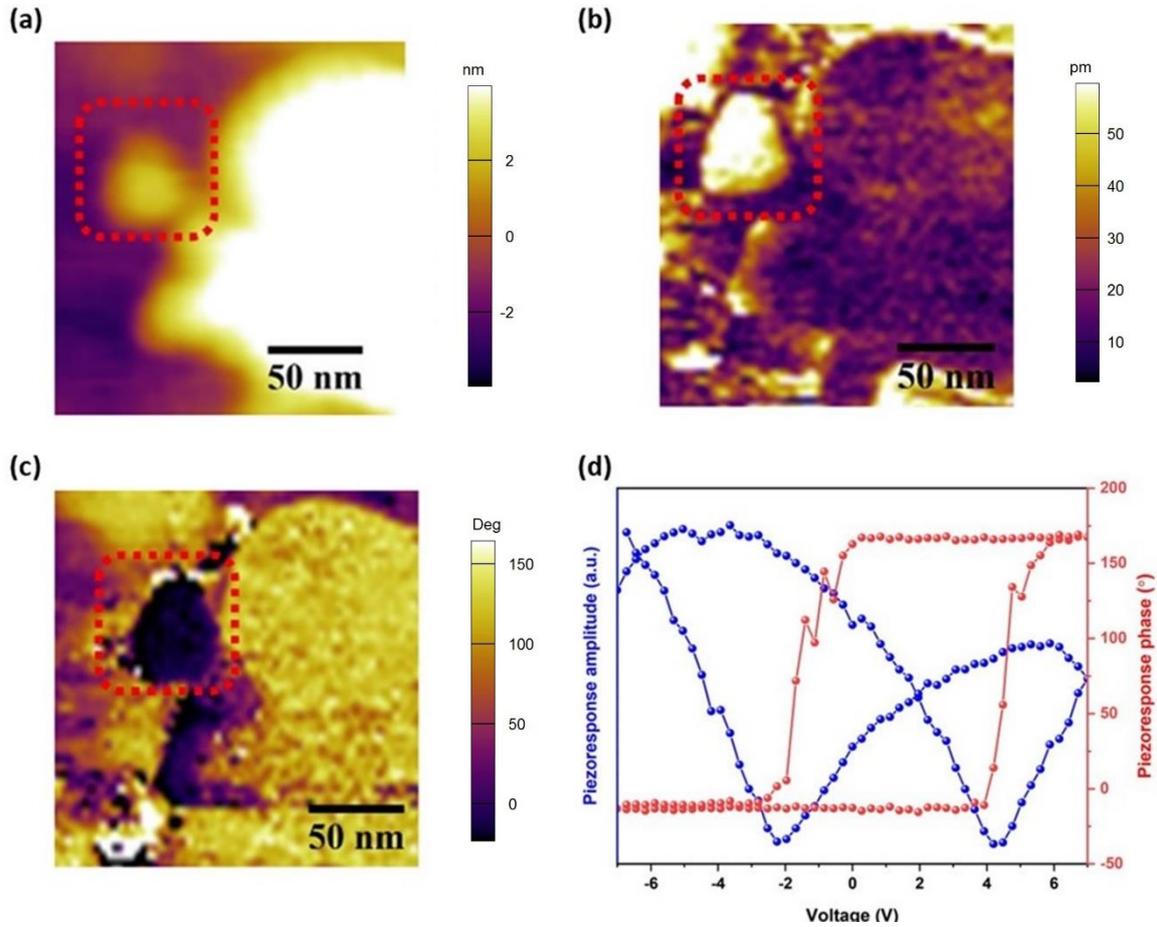

**Figure 6. PFM results of defective BTO-H nanoparticles:** (a) Topography, (b) Amplitude and (c) Phase images of a BTO-H nanoparticle, (d) Amplitude-voltage butterfly and phase-voltage hysteresis loops.

**Figure 7(a)** shows the extent of RhB photodegradation, where the degradation increases with increasing irradiation time and that all catalysts show photoactivity under UV light. These curves can be fit to the generally observed pseudo-first-order kinetics, which are described by **Equation 12** [69]:

$$\ln\left(\frac{C}{C_0}\right) = -kt \tag{12}$$

where:

$c_0$ = Initial concentration of RhB solution
$c$ = Final concentration of RhB solution
$k$ = Rate constant
$t$ = Time

The catalytic performance with UV irradiation only is in the order BTO-O < BTO-N < BTO-A < BTO-H and the extent of catalytic degradation for BTO-H was ~99% in 60 mins.

(18)

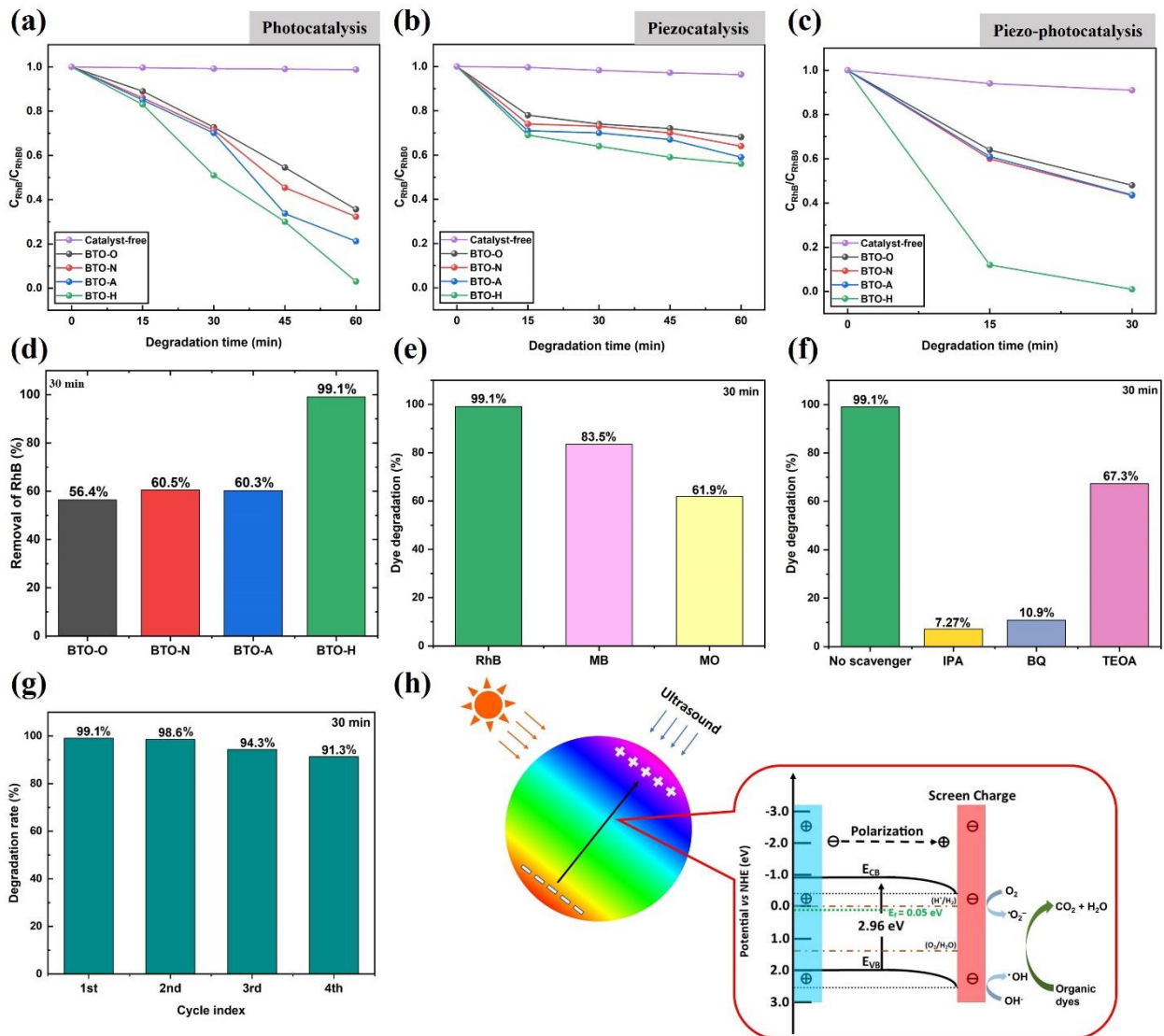

**Figure 7.** Performance of defective BTO nanoparticles following 10 min sonication: (a) Photo-degradation of RhB as a function of time, (b) Piezo-degradation of RhB as a function of time, (c) Piezo-photo-degradation of RhB as a function of time, (d) Extent of piezo-photo-degradation of RhB solution after UV irradiation for 30 min, BTO-H: (e) Extent of piezo-Piezo-photo-degradation of RhB, MB, and MO solutions after UV irradiation for 30 min, (f) Piezo-photodegradation efficiency of RhB solution with different quenchers after UV irradiation for 30 min (IPA = isopropyl alcohol, BQ = P-benzoquinone, TEOA = triethanolamine), (g) Reusability testing after piezo-photo-degradation for four cycles (30 min/cycle), (h) Conceptual energy band diagram for charge separation and transfer in BTO-H under irradiation and sonication

**Figure 7(b)** reveals that the extents of RhB degradation with only sonication (40 kHz, 100 W) exhibit a consistent trend with photodegradation, although a maximum of only ~55% degradation ratio occurs for BTO-H in 60 min. **Figure 7(c)** reveals that the extents of RhB degradation with both UV irradiation and sonication (40 kHz, 100W) exhibit differential trends, where the degradation of BTO-H reached >99% in 30 min but the less defective samples reached a maximum of only ~61%. However, all four of these performances are superior to those of UV irradiation or sonication alone. **Figure 7(d)** emphasizes that the RhB degradation at the fixed time point of 30 min is considerably superior for BTO-H. There are many materials

(19)

factors that may affect these data, including phase assemblage, impurities (*viz.*, heterojunction formation), crystallinity, crystallite size, electronic band alignment, recombination time, BET surface area, [$V_O^{\bullet\bullet}$], and blockage of active sites [70].  Accordingly, the present work suggests that the catalytic performance (dye degradation and related water splitting) is associated with the following:

- The reduced $E_g$ and shift in $E_f$, which are caused by increased defect concentrations ($Ti^{3+}$, $Ti^{2+}$, $Ti^0$, $V_O^{\bullet\bullet}$) and optimized by $H_2$ reduction, result in band alignment that establishes conditions favorable for dye degradation and related water splitting.
- The shallow energy level established by the high $E_f$ and associated lowering of the CBM enhances charge carrier separation by promoting the diffusion of electrons to the surface, thereby reducing recombination with holes and facilitating higher carrier concentrations and efficient charge transfer across the grain surface.
- The increased BET surface area and associated [$V_O^{\bullet\bullet}$] are optimized by $H_2$ reduction, thereby providing more active sites for catalytic reaction, although there may be mitigation by blockage of active sites by species dissolved in the water.
- The application of mechanical vibration leads to displacement of charges (and possibly adsorbed species), thereby inducing the piezoelectric potential that facilitates surface reactions.
- The sonication and resultant external stress induce the reduced band gap and band bending that shifts the conduction band closer to the HER band potential and the valence band further away from the OER band potential, thus enabling the former to increase $H_2$ generation from water splitting and allowing the latter to decrease $O_2$ generation.

**Figure 7(e)** reveals that equivalent testing using MB and MO dyes also showed substantial extents of degradation within the same time frame of 30 min.  These dyes were selected for comparative testing owing to differences in the wavelengths of maximal visible light absorption relative to that of RhB.  That is, RhB absorbs maximally in the green range at 554 nm while MB absorbs most strongly in the red range at 664 nm and MO absorbs most strongly in the blue range at 464 nm [70].  These data suggest that absorption may not be a serious limitation to the decomposition of a wide spectrum of organic contaminants of water purified by piezo-photocatalysis.  **Figure 7(f)** shows the roles of different ROS in the mechanisms of RhB degradation during piezo-photocatalysis, as revealed by the effects of the scavengers triethanolamine (TEOA), P-benzoquinone (BQ), and isopropyl alcohol (IPA) [71].  These three scavengers are known to quench active holes ($h^+$), the superoxide radical ($^\bullet O_2^-$), and the hydroxyl radical ($^\bullet OH$), respectively.  These data show that the order of effectiveness of these ROS in piezo-photodegradation is TEOA < BQ < IPA.  These results indicate that $^\bullet O_2^-$ and $^\bullet OH$ are the principal ROS that contribute to the degradation of RhB.  **Figure 7(g)** shows the stability and recyclability data for BTO-H in terms of four-cycle testing and the reduction in the degradation extent.  The good cycling stability during piezo-photodegradation is similar to that of other studies on BTO done using both piezocatalysis [41,72] and piezo-photocatalysis [7,25].  Thus, the imposition of a high defect density did not destabilize the material to the point of making it susceptible to leaching.

**Figure 7(h)** illustrates a schematic of the energy band diagram for the piezo-photocatalytic degradation of organic phases using defective BTO subjected to vibration and irradiation.  The generation of the ROS ($^\bullet O_2^-$ and $^\bullet OH$) and associated electron-hole separation confirm that the catalytic performance is enhanced through the combination of the five factors described above.  Photocatalysis reactions are initiated by excitation from radiation energy ($h\upsilon$) equal to or greater than the $E_g$.  As a result, photogenerated electrons ($e^-$) are promoted from the VB to the CB, leaving a hole ($h^+$) in the VB. However, the application of ultrasonic stress can enhance the rapid generation of electrons further owing to two mechanisms [73]. First, electrons are generated from the high pressures (~$10^2$ MPa) caused by the collapse of ultrasound-induced cavitation bubbles in the solution. These pressures are sufficient to generate localized external temperatures as high as ~5000°C on the grain surface, thereby readily exciting free electrons at the catalyst-water interface. Second, electrons are generated in the bulk owing to the general internal thermal excitation induced by vibration.

The effectiveness of the electrons can be heightened by the strong piezoelectric effect of the BTO, which establishes an internal electrical field that facilitates the migration of excited electrons and holes to opposite surfaces, thereby establishing a surface potential that encourages the adsorption of dissociated ions, such as $H^+$ and $OH^-$, which can facilitate both ROS production and the water splitting [73]. Further, the accumulation of piezoelectrically-induced charges on the opposite material surfaces alters the density of states, thus resulting in bending of the surface band edges and establishing favorable band edges for the degradation of organic phases and water splitting. In particular, under a static deformation, the piezopotential drives the opposing mobile charges to continually accumulate on the two polar side to create a depolarization field which counters the piezoelectric polarization effect and thus screen band bending [73]. Therefore, an alternating lattice strain is required, such as that produced by ultrasound, as the accompanied variation in the piezoelectric field enables the internal mobile charges to be in a metastable state rather than accumulating in an oriented manner, thereby preventing the band bending screen. While these concepts are relevant to stoichiometric BTO, the presence of oxygen vacancies in defective BTO further enhances catalysis through both adsorption of $O_2$ and capture of free electrons, which promoted the activation of $O_2$ to $^{\bullet}O_2^-$. The present work and that of others [24,74,75] shows that $V_O^{\bullet\bullet}$ play a critical role in enhancing the preceding processes. Wang et al. [24] investigated the [$V_O^{\bullet\bullet}$] in BTO and shows that increasing [$V_O^{\bullet\bullet}$] could efficiently adsorb and active $O_2$ on the surface of BTO and consequently enhance piezocatalytic activity, which is consistent with present work. Liu et al. [74] regulated the [$V_O^{\bullet\bullet}$] in $Na_{0.5}Bi_{0.5}TiO_3$ by both experiment and simulation, showing that increasing [$V_O^{\bullet\bullet}$] provides more free surface charges and facilitates charge transfer, thereby improving the piezocatalytic activity. The increasing [$V_O^{\bullet\bullet}$] also promotes the adsorption of $^{\bullet}O_2^-$ and $^{\bullet}OH$ and also activates the O-O bond and O-H bond to generate more ROS. Li et al. [75] examined the piezo-photocatalytic performance of $AgNbO_3$ and demonstrated the principal positive effect of $V_O^{\bullet\bullet}$. This was attributed to the role of $V_O^{\bullet\bullet}$ in domain switching, facilitating the inversion at lower electric fields. Increasing [$V_O^{\bullet\bullet}$] also was observed to narrow the $E_g$, improve optical absorbance, and, most importantly, promote the separation of electron-hole pairs under an external electric field.

The $H_2$ evolution rates by piezocatalytic and piezo-photocatalytic activation of BTO also were investigated in the absence of sacrificial agent or co-catalyst. These data, which are for DI water and, for the first time, simulated and natural seawater, are shown in **Figures 8(a)** and **(b)**. These data, which are for piezocatalysis for 2 h followed by piezo-photocatalysis for 3 h, reveal that the rates follow the order: BTO-O < BTO-N < BTO-A < BTO-H. This trend is consistent with the dye degradation data in **Figures 7(a-d)**, again confirming the key role of the increasing defect concentrations ($Ti^{3+}$, $Ti^{2+}$, $Ti^0$, $V_O^{\bullet\bullet}$) and maximization by $H_2$ reduction. These data show that the $H_2$ evolution rates exhibit approximately linear trends, demonstrating that, at least over the time frame of 5 h, the $H_2$ production is constant, with a maximal rate of 100.7 µmol/g/h (BTO-H), which is comparable to the rates reported by You et al. [29] for $BiFeO_3$ (124.1 µmol/g/h) and Su et al. [25] for $BaTiO_3$ (159 µmol/g/h).

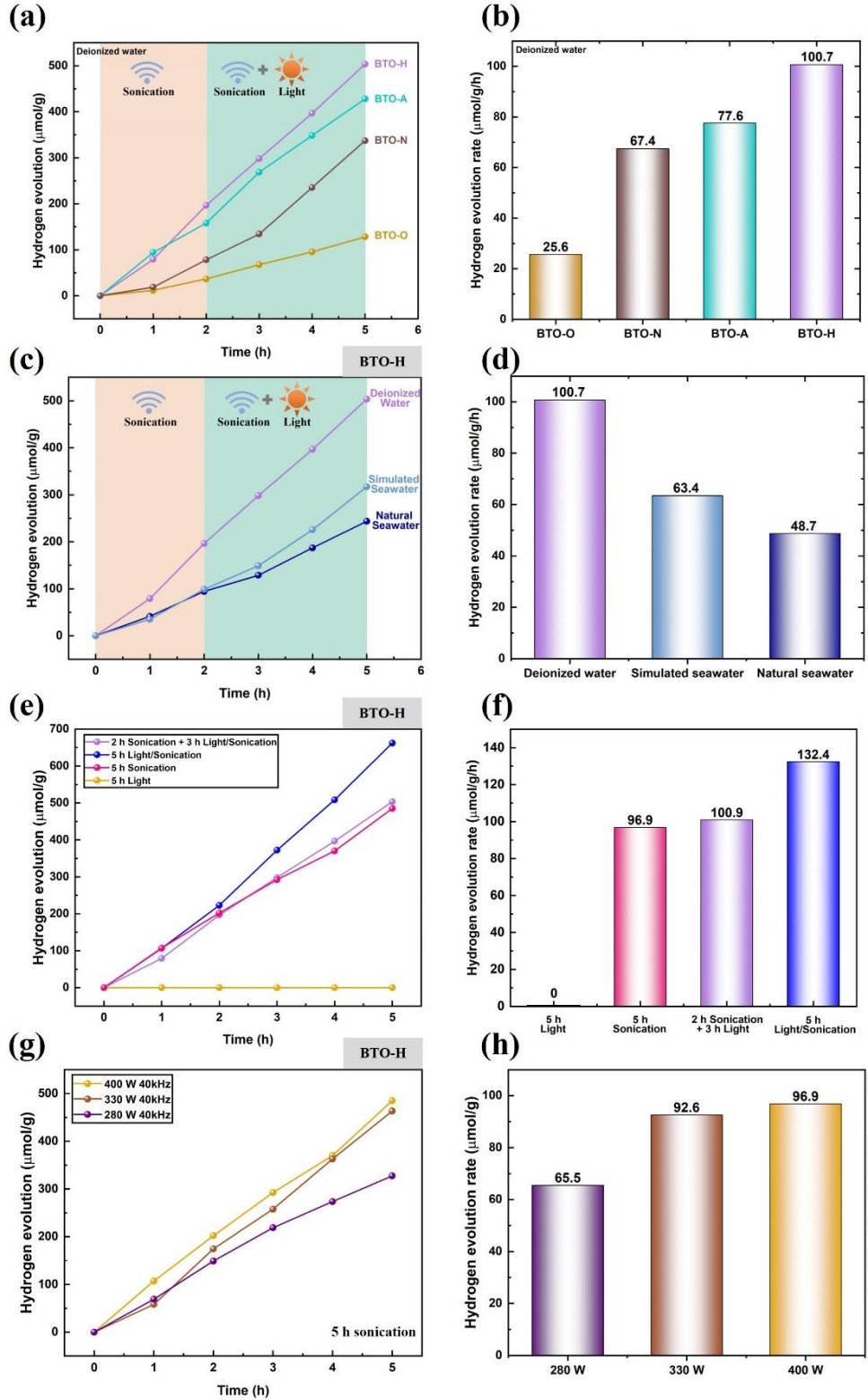

**Figure 8. Effects of test conditions on H$_2$ evolution by defective BTO:** (a) Data for deionized water, (b) Rates for deionized water, (c) Data for different waters, (d) Rates for different waters, (e) Data for different activation methods, (f) Rates for different activation methods, (g) Data for different piezocatalysis power levels, (h) Rates for different piezocatalysis power levels

**Figures 8(c)** and **(d)** show the effect of water type (DI water, simulated seawater and natural seawater) for BTO-H, revealing again that the trends of the H$_2$ evolution rates were linear.

(22)

However, the rate from DI water was significantly greater than those of the seawaters. These data are consistent with the effects of the number of types and concentrations of dissolved ions. Whereas simulated seawater contains only dissolved NaCl at a standard concentration of 3.5 wt% (plus K contaminants), the natural seawater acquired from Coogee Beach, Sydney, Australia was of similar salinity but with greater contamination level[1]. It also contains suspended solids as well as dissolved inorganic and organic carbon. These data are given in **Table 1**, which provides the relevant concentrations before and after piezo-photo-catalytic $H_2$ evolution. In light of Na adsorption and retention on the nanoparticles, these data suggest that NaCl is the main factor behind these differences, where $Na^+$ would cause electron ($e^-$) scavenging and $Cl^-$ would cause hole ($h^+$) scavenging. The difference between simulated and natural seawater is attributed to the greater concentrations of $K^+$, $Mg^{2+}$, $Ca^{2+}$, and $SO_4^{2-}$ [10,21]. Ji *et al.* [17] showed that $Mg^{2+}$ can reduce significantly the $H_2$ evolution. These findings were supported by those of Li *et al.* [76], who showed that the chelation of $Mg^{2+}$ blocks the separation of photogenerated electron-hole excitons and induce charge redistribution, thereby negatively impacting on the photocatalytic performance. Ji *et al.* [17] also showed that the presence of sulfite and sulfide ions, which act as sacrificial agents, reduce the $H_2$ evolution as these ions can cause charge imbalance and resultant irregular ROS generation in the solution, thereby impeding their function as sacrificial agents by forming precipitates [17]. Another factor is the blockage of surface active sites by these ions as well as by suspended and dissolved species [10,21]. Solids suspended in seawater decrease light absorption by scattering and dissolved species may alter the refractive index and so increase reflection. These effects also can provide the energy to trigger biochemical reactions that can reduce the $H_2$ production as well [77].

**Table 1** also shows that the dissolution of NaCl shifted the pH from slightly basic to notably acidic, which might result from the formation of carbonic acid [77]. **Table 1** also shows that $H_2$ evolution resulted in consistent increases in the pH for all three types of water. The resultant decreases in [$H^+$] would be likely to assist $H_2$ evolution owing to LeChatelier's principle [78], where the chemical gradient would enhance $H^+$ dissolution.

**Table 1** also reveals that [Na] is less than [Cl], although they should be the same. This apparent anomaly was examined by multiple washing and XPS analysis of the adsorbates on the nanoparticles. These data are given in **Table 2** and they reveal that [Na] is consistently greater than [Cl], which is the converse for the data for the corresponding solutions. Consequently, it is clear that Na is retained more readily as an adsorbate while the Cl is more readily removed by washing, which explains these two sets of oppositely differential data. This suggests that, if blockage of surface active sites is relevant, then the reduction in $H_2$ evolution is a result of the effect of $Na^+$ rather than $Cl^-$. This is consistent with the zeta potential of $BaTiO_3$, which is negative at almost all pH values [79]. These results suggest that the process, kinetics, and economics of seawater splitting can be manipulated through dilution, thereby reducing the effects of the dissolved species (and suspended solids), and the pH control, thereby increasing the $H^+$ solubility.

---

[1] Assessment of the salinity by [Na] is misleading owing to XPS data showing enhanced Na concentration on the surfaces of the separated nanoparticles. Assessment of the salinity by [Cl] should accommodate the assumption of the dissolution of K, Mg, and Ca chlorides, which results in [Cl] values within ~2% for simulated and natural seawaters.

Table 1. Summary of pH and ICP-OES elemental analyses of different waters before and after piezo-photo-catalytic $H_2$ evolution

| Water Type | Before or After HER | pH | Na mg/L (ppm) | Cl mg/L (ppm) | K mg/L (ppm) | Mg mg/L (ppm) | Ca mg/L (ppm) | S mg/L (ppm) | C mg/L (ppm) |
|---|---|---|---|---|---|---|---|---|---|
| Deionized Water | Before | 7.44 | 0.00 | 0.00 | 0.00 | 0.00 | 0.00 | 0.00 | 0.00 |
| Simulated Seawater | Before | 5.30 | 11,021 | 15,916 | 7.00 | 0.10 | 0.79 | 0.05 | 0.00 |
| Natural Seawater | Before | 5.84 | 11,187 | 18,900 | 396.80 | 1,496.00 | 441.00 | 925.00 | 24.27 |
| Deionized Water | After | 8.38 | 0.26 | 0.16 | 0.00 | 0.00 | 0.00 | 0.00 | 0.00 |
| Simulated Seawater | After | 6.04 | 11,084 | 19,958 | 10.10 | 0.08 | 0.47 | 0.86 | 5.55 |
| Natural Seawater | After | 6.20 | 11,532 | 23,159 | 399.30 | 1,493.00 | 433.00 | 925.99 | 25.77 |

**Table 2.** Summary of XPS elemental analyses of different powders after piezo-photocatalytic HER in simulated seawater or natural seawater

| Seawater | Washing | Ba (at%) | Ti (at%) | Na (at%) | Cl (at%) |
|---|---|---|---|---|---|
| Simulated | 10 mins (15 mL) | 10.82 | 12.66 | 41.53 | 34.98 |
| | 40 mins (40 mL) | 9.31 | 10.09 | 51.29 | 29.30 |
| Natural | 10 mins (15 mL) | 31.04 | 16.99 | 41.18 | 10.78 |
| | 40 mins (40 mL) | 28.61 | 39.21 | 19.16 | 13.01 |

**Figures 8(e)** and **(f)** show the effects of the activation methods for BTO-H, revealing the following features about the linear $H_2$ evolution rates:

- Photocatalysis of BTO is inactive, which is attributed to the very rapid rate of electron-hole recombination observed in this material [80].
- Piezocatalysis of BTO is effective, giving $H_2$ evolution rates comparable to those of other BTO samples [7,25] and other materials, including $BiFeO_3$ [29], ZnS [81], $MoS_2$ [82], etc. This is attributed to the following factors:
  (1) The electrons generated from high pressures caused by the collapse of ultrasound-induced cavitation bubbles generate localized extremely high temperatures on the grain surface and so excite free charges to diffuse to the surface.
  (2) The electrons generated in the bulk from vibration-induced thermal excitation are given sufficient energy to bridge the $E_g$.
  (3) The vibration-induced lowering of the $E_g$ from the ultrasound also facilitates the capacity of the electrons to bridge the $E_g$.
  (4) The high electron densities and rapid transfer rates deriving from (2) and (3) result in low electron-hole recombination rates.
  (5) The electron accumulation deriving from (2) and (3) establish the band bending that facilitates the HER while suppressing the OER, as discussed subsequently.
- Piezo-photocatalysis is more effective giving an $H_2$ evolution rate ~37% greater than that of piezocatalysis alone. This is attributed to the high electron densities, rapid transfer rates, and resultant low electron-hole recombination rates in (4), which overcome the hindrance to photocatalysis in (1). Consequently, piezocatalysis and photocatalysis provide synergistic effects in piezo-photocatalysis.

**Figure S8**, which reports the $O_2/N_2$ molar ratio, provides indicative data for the effects of the activation methods on the OER. These data are tentative owing to the limited accuracy of $O_2$ volume measurement. Since the OER data in **Figure S8** are consistent with the HER data in **Figures 8(e)** and **(f)**, then this supports the conclusion that the OER data are relatively reliable.

Also, the enhancement of the HER in **Figures 8(e)** and **(f)** and the suppression of the OER in **Figure S8** are supported by the band bending indicated in the DFT simulations in **Figure 5(b)**. However, since OER occurred with piezo-photocatalysis, then this indicates that the band bending is relatively small. However, the HER can remain effective owing to the proximity of the $E_f$ to the CB.

**Figures 8(g)** and **(h)** show the effects of ultrasound power alone on the H$_2$ evolution rates for BTO-H, which are linear. These data reveal a logarithmic trend for latter as a function of the former. The governing relation between the power, which is a reflection of the external stress, and the HER, which is a reflection of the piezoelectric charge, is given by **Equation 13** [29], which is linear:

$$Q = d \cdot \sigma \qquad (13)$$

where:

Q = Piezoelectric charge per unit area
d = Piezoelectric coefficient
σ = External stress

The logarithmic trend of the data in **Figure 8(h)** is likely to reflect the effectiveness of the power level in deagglomeration, thereby exposing more surface active sites for piezocatalysis, where 280 W was less effective than the two higher power levels, which exhibited similar H$_2$ evolution rates.

**Figure 9** applies the preceding body of data to conceptualize the photocatalytic, piezocatalytic, and photo-piezocatalytic H$_2$ evolution reaction mechanisms:

- **Photocatalysis:** With defective BTO under visible light irradiation, electrons bridge the $E_g$ and reach the CB, thus forming corresponding holes in the VB. However, these charge carriers rapidly recombine, thereby preventing HER and OER.
- **Piezocatalysis:** Defective BTO subjected to ultrasound alone benefits from the mechanical force directly excites the electrons to bridge the $E_g$ owing to electron generation from bubble cavitation and thermal excitation, lowering of the $E_g$ from vibration, resultant high electron densities and rapid transfer rates, and significant band bending. The net electric field from the sum of the self-polarization and the stress-induced polarization, selectively attracts the electrons and holes, both of which accumulate and thus lead to band bending. Ultimately, this band bending facilitates the HER while suppressing the OER. In the absence of OER, change balance is maintained by the retention of OH$^-$ ions from the splitting of H$_2$O, which raises the pH. Simultaneously, the tendency for the unstable surface Ti$^{2+}$ and Ti$^{3+}$ ions to oxidize is suppressed by the high electron densities.
- **Piezo-photocatalysis:** Defective BTO subjected to visible light irradiation and ultrasound together experiences benefits similar to those of piezocatalysis except that the process is enhanced through the contribution of photocatalysis. The inactivity of BTO to photocatalysis is overcome by the high electron densities, rapid transfer rates, and low electron-hole recombination rates established by piezocatalysis. However, the charge carrier densities are reduced as some of these are utilized in the photocatalysis. In this case, the extent of band bending is reduced, as evidenced by the occurrence of OER with piezo-photocatalysis. Fewer OH$^-$ ions are produced as some of the H$_2$O is utilized in the OER, so the pH would be expected to increase less.

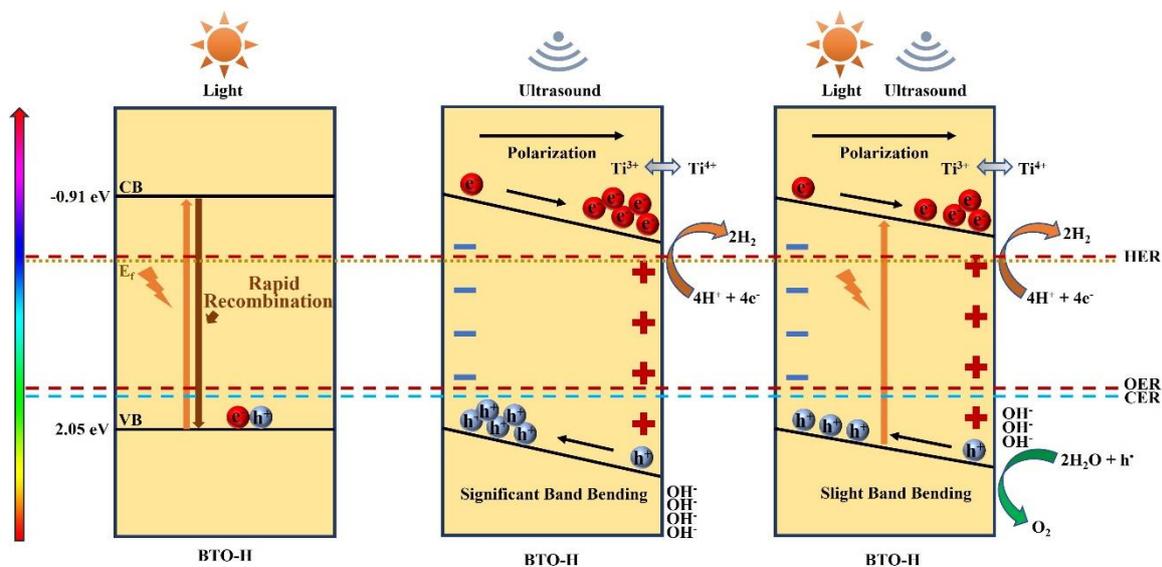

**Figure 9.** Schematic illustrations of band bending of defective BTO-H nanoparticles as a function of activation method

## 3. Summary and Conclusions

In present work, cubic BTO was converted to defective BTO nanoparticles by heat-treatment under $O_2$, $N_2$, Ar, or $H_2$ at 800°C for 12 h in order to vary the extent of reduction. The effect of reduction on the crystallography was examined by XRD, Raman, SEM-EDS, and TEM-SAED. These data revealed that reduction caused the partial conversion of cubic BTO to tetragonal. TEM, SEM and laser diffraction data revealed monodisperse cuboidal morphologies of typically 30-60 nm size after deagglomeration by 30 min sonication. The crystallographic data were confirmed by HAADF-STEM images, which showed inhomogeneously distributed darker domains of lower true density that corresponded to tetragonal BTO. This expanded structure was stabilized by the expansion resulting from the $Ti^{4+} \rightarrow Ti^{3+} \rightarrow Ti^{2+}$ multiple reductions, which were charge-compensated by oxygen vacancy formation. These valence changes were confirmed by EPR, XPS, TEM-EELS and ARXPS data. The oxygen vacancy concentrations were determined by XPS and these correlated solely with the extent of BET surface area.

The trends in the TEM-EELS and ARXPS data indicate the expected observation of increasing reduction with increasing depth. The quantitative ARXPS data suggest that the Ti valence concentration gradient across the surface layer (~1 nm) is linear; the linearity of the $Ti^{4+}$ concentration suggests with less certainty a similarly linear trend in the Ti valence concentration across the subsurface layer (~1 nm). Critically, these data show for the first time that there is a discrete and identifiable interface between the surface and subsurface, which is revealed by the inflection in the $Ti^{(4-x)+}$ concentration gradients. Further, the extrapolation of the ARXPS data to 0° allows the identification of the relative proportions of the three Ti valences at the absolute surface.

The EPR, XPS, TEM-EELS, and ARXPS reveal information about the presence of different ionic defects. However, the PL data suggest that it is possible to differentiate between the presence of ionic defects and electronic defects. The latter are associated with all three color centers deriving from oxygen vacancies ($F^0$, $F^+$, $F^{++}$). These combined data represent a starting point to propose the relevant defect equilibria, which confirm the natures of the defects and their mechanisms of charge compensation, which is redox-ionic, thereby correlating with the observation of reduced Ti valences and charge-compensating oxygen vacancy formation.

The full electronic band structures were determined on the basis of XPS-VB, UV-Vis Reflectance, and AM-KPFM data. These data revealed that the significant $H_2$ reduction resulted in a partly decrease in the $E_g$ and a shift in the $E_f$ to nearly the HER potential. The changes in these two parameters strongly suggest that $H_2$ reduction of BTO facilitates the HER and hence the possibility of splitting water more effectively. The DFT simulations suggest the effects of ultrasound on the electronic band structure, where the cyclic amplitude switching of the ultrasound results in cyclic compressive-tensile strains. However, the effects of these on the potential change are differential. Compression significantly lowers the $E_g$ while tension slightly increases it. The CB is shifted toward the HER under both compression and tension while the different trends in the VB shifts under compression and tension result in an overall shift away from the OER from the associated band bending. Consequently, this band bending, which is caused by the self-polarization of the BTO, can facilitate the splitting of water using ultrasound. The directions of band bending are consistent with more efficient generation of hydrogen while suppressing the generation of oxygen. The PFM data confirm the strong piezoelectric performance as well as significant self-polarization, which derives from the high oxygen vacancy concentration deriving from the $H_2$ reduction.

The degradation of multiple dyes and the application of photocatalysis, piezocatalysis and piezo-photocatalysis revealed that $H_2$ reduction of BTO facilitated effective RhB degradation of 99% in 30 min. More importantly, this highly defective BTO exhibited a null results with photocatalysis but a very efficient $H_2$ evolution rate of 96.9 μmol/g/h with piezocatalysis and a highly efficient $H_2$ evolution rate of 132.4 μmol/g/h with piezo-photocatalysis. These data demonstrate that, although photocatalytic water splitting alone had no effect, in combination with ultrasound, it contributed to the overall effect. This was attributed principally to the electrons generated from high pressures resulting from the collapse of ultrasound-induced cavitation bubbles. The high pressures generated extremely high temperatures on localized regions of the grain surface, resulting in high electron densities, rapid transfer rates, and resultant low electron-hole recombination rates, all of which contribute to more efficient water splitting.

The comparative $H_2$ evolution rates as a function of water type were 100.7 μmol/g/h for DI water, 63.4 μmol/g/h for simulated seawater, and 48.7 μmol/g/h for natural seawater. These differences were attributed largely to the role of the dissolved ions in seawater through their effects in charge-carrier scavenging, prevention of charge separation, solution charge imbalance and resultant irregular ROS generation, blockage of active sites by suspended and dissolved species, and/or triggered biochemical reactions. The concentrations of these ions were confirmed by ICP-OES of the solutions and XPS of the solids. These data suggest that the process, kinetics, and economics of seawater splitting can be manipulated through dilution, thereby reducing the effects of the dissolved species (and suspended solids), and the pH control, thereby increasing the $H^+$ solubility.

The present work discloses the initial report of piezo-photocatalysis of seawater splitting. These experimental and simulation data suggest that $H_2$ reduction of BTO result in an energy band structure and band bending facilitate efficient $H_2$ generation while contributing to the suppression of simultaneous $O_2$ generation. This approach of piezo-photocatalysis represents a new strategy for commercial $H_2$ production, potentially using large-scale green technologies, including ultrasound vibrations deriving from wind energy and/or tidal energy (piezocatalysis) and sunlight (photocatalysis). Critically, these technologies can overcome the limited supplies of freshwater, the necessity of its purification, and make full utilisation of abundant seawater and sunlight as well as geotechnical natural resources (large land masses, extensive shorelines, regular tidal patterns, and irregular wind patterns). These technologies have the potential to provide the bases for novel systems aimed at a net zero emissions future while driving economic growth.

**Conflicts of Interest**

There are no conflicts to declare.

**Acknowledgements**

The authors acknowledge the financial and infrastructural support as follows:

- Tuition Fee Scholarship from UNSW Sydney and supplementary support from the China Scholarship Council.
- Electron Microscope Unit, Mark Wainwright Analytical Centre, UNSW Sydney
- Spanish Ministry of Science, Innovation and Universities under the "Ramon y Cajal" fellowship RYC2018-024947-I.
- Microscopy Australia (formerly known as AMMRF) and the Australian National Fabrication Facility (ANFF) at Flinders University

The authors also wish to acknowledge the technical assistance of Dr. Dawei Zhang and Mr. Haotian Wen, School of Materials Science and Engineering, UNSW Sydney, for consultations and assistance with PFM and HAADF-STEM analysis.

ESCA including application to organometallics. J. Electron Spectrosc. Relat. Phenom.,1996, **77**(1), 41-57.